\documentclass[10pt,aps,prd,twocolumn,superscriptaddress,amsmath, nofootinbib]{revtex4-1}

\usepackage{capt-of}
\usepackage{placeins}
\usepackage{aas_macros}
\usepackage{lipsum}
\usepackage{tikz}
\usepackage{layouts}
\usepackage{colortbl}
\usepackage[hidelinks]{hyperref}
\usepackage[capitalize]{cleveref}
\Crefname{equation}{Equation}{Equations}
\crefname{equation}{Eq.}{Eqs.}
\Crefname{figure}{Figure}{Figures}
\crefname{figure}{Fig.}{Figs.}
\Crefname{table}{Table}{Tables}
\crefname{table}{Tab.}{Tabs.}
\Crefname{section}{Section}{Sections}
\crefname{section}{Sec.}{Secs.}

\usepackage{amsthm}
\theoremstyle{definition}
\newtheorem{proposal}{Proposition}
\crefname{proposal}{Proposition}{Proposition}

\newcommand{\Planck}{\textit{Planck}}
\newcommand{\DES}{DES}
\newcommand{\SDSS}{BOSS}
\newcommand{\SHOES}{S$H_0$ES}
\newcommand{\CosmoChord}{\texttt{CosmoChord}}
\newcommand{\PolyChord}{\texttt{PolyChord}}
\newcommand{\MultiNest}{\texttt{MultiNest}}
\newcommand{\anesthetic}{\texttt{anesthetic}}
\newcommand{\CosmoMC}{\texttt{CosmoMC}}
\renewcommand{\d}[2][]{\operatorname{d}^{#1}\!{#2}}

\begin{document}
\title{Quantifying tensions in cosmological parameters:\\Interpreting the \DES{} evidence ratio}
\author{Will Handley}
\email[]{wh260@mrao.cam.ac.uk}
\affiliation{Astrophysics Group, Cavendish Laboratory, J.J.Thomson Avenue, Cambridge, CB3 0HE, UK}
\affiliation{Kavli Institute for Cosmology, Madingley Road, Cambridge, CB3 0HA, UK}
\affiliation{Gonville \& Caius College, Trinity Street, Cambridge, CB2 1TA, UK}

\author{Pablo Lemos}
\email[]{pablo.lemos.18@ucl.ac.uk}
\affiliation{Department of Physics and Astronomy, University College London, Gower Street, London, WC1E 6BT, UK}

\begin{abstract}
    We provide a new interpretation for the Bayes factor combination used in the Dark Energy Survey (DES) first year analysis to quantify the tension between the DES and \Planck{} datasets. The ratio quantifies a Bayesian confidence in our ability to combine the datasets. This interpretation is prior-dependent, with wider prior widths boosting the confidence. We therefore propose that if there are any reasonable priors which reduce the confidence to below unity, then we cannot assert that the datasets are compatible. Computing the evidence ratios for the \DES{} first year analysis and \Planck{}, given that narrower priors drop the confidence to below unity, we conclude that \DES{} and \Planck{} are, in a Bayesian sense, incompatible under $\Lambda$CDM\@. Additionally we compute ratios which confirm the consensus that measurements of the acoustic scale by the Baryon Oscillation Spectroscopic Survey (\SDSS{}) are compatible with \Planck{}, whilst direct measurements of the acceleration rate of the Universe by the \SHOES{} collaboration\footnote{Supernovae and $H_0$ for the Equation of State.} are not. We propose a modification to the Bayes ratio which removes the prior dependency using Kullback-Leibler divergences, and using this statistical test find \Planck{} in strong tension with \SHOES, in moderate tension with \DES{}, and in no tension with \SDSS{}. We propose this statistic as the optimal way to compare datasets, ahead of the next DES data releases, as well as future surveys. Finally, as an element of these calculations, we introduce in a cosmological setting the Bayesian model dimensionality, which is a parameterisation-independent measure of the number of parameters that a given dataset constrains.
\end{abstract}

\maketitle

\section{Introduction}

The analysis of the first year of data from the Dark Energy Survey~\cite{DESParameters2017} (henceforth DES Y1) has generated considerable discussion. DES Y1 analysed data from cosmic shear, galaxy clustering, and galaxy-galaxy lensing (an analysis they refer to as ``3x2'' since it combines three two-point functions). This data combination is particularly suited to constraining the present day matter density $\Omega_m$ and the parameter $\sigma_8$, defined as the present-day linear theory root-mean-square amplitude of the power spectrum of matter fluctuations, averaged in spheres of radius $8 h^{-1} {\rm Mpc}$, where $h$ is the Hubble constant in units of $100~{\rm km~s^{-1}~Mpc^{-1}}$. Before the publication of DES Y1, this parameter combination measured by weak lensing had already generated controversy, with claims of tensions with respect to the Cosmic Microwave Background (CMB) values measured by \Planck{}~\cite{PlanckParameters2018} by both the CFHTLenS and Kilo Degree Survey (KiDS) collaborations~\cite{Joudaki2017, Kohlinger2017, Hildebrandt2017}. Whilst this discrepancy has led to claims of new physics \cite{Joudaki2017b}, it has also highlighted unknown problems in weak lensing analyses that have reduced these tensions to below significant levels \cite{Efstathiou2018, Troxel2018, Kohlinger2019}. 

DES Y1 obtained results that appear to be in mild tension with \Planck{} (see Fig. 10 of DES Y1), but are reported to be perfectly consistent according to the evidence ratio statistic\footnote{Here $R$ refers to the Bayes factor combination used in DES Y1 to compare different datasets, not to the Bayes ratio used to compare models.} $R$ used in their analysis to quantify the degree of discordance between 3x2 and CMB data. Whilst this $R$ statistic was proposed some time ago~\cite{Marshall2006}, and supported since then by many cosmologists~\cite{Trotta2008,Verde2013,Verde2014,Raveri2016a,Seerhars2016}, it is particularly relevant to consider its precise interpretation in light of present and future tensions arising with increasingly powerful datasets providing ever more precise parameter constraints. Other measures of tension between datasets have been proposed in the past~\cite{Inman1989,Battye2015,Seehars2014,Nicola2019,Kunk2006,Karpenka2015,MacCrann2015,Adhikari2019,,Douspis,Raveri2018}. A summary of a lot of these methods can be found in \cite{Charnock:2017}. 

In this paper we argue that $R$ is an appropriate measure of tension, quantifying the Bayesian degree of confidence in the ability to combine the data. However, $R$ has some subtle prior-dependent properties, which has led to its misuse in previous works. We explain these properties and provide Bayesian methods to correctly calibrate the scale on which it sits. We also propose an alternative statistic that preserves the desired properties of $R$ to compare different datasets, including its Bayesian nature, but does not suffer from undesired prior dependences.

The tension between weak galaxy lensing and \Planck{} is not the only existing tension in cosmology. Measurements of the expansion rate of the Universe parameterised by the Hubble constant $H_0$ using Type Ia supernovae calibrated by the period-luminosity relation of Cepheids and local distance anchors by the \SHOES{} collaboration \cite{Riess2016, Riess2018} are in tension with the \Planck{} value inferred from the CMB using a $\Lambda {\rm CDM}$ cosmology~\cite{PlanckParameters2018}. We use this case as an example of clear tension between experiments. Conversely, the measurements of the Baryon Acoustic Oscillation (BAO) scale and Redshift-Space Distortions (RSD)  by BOSS \cite{SDSS} produce values of the parameters $\Omega_m$ and $\sigma_8$ that are in good agreement with \Planck{}. We use this case as an example of no tension between experiments. 

 
The paper is structured as follows:
In \cref{sec:background} we briefly review the key Bayesian theory and establish notation.
In \cref{sec:R} we define the logarithmic Bayes and information ratios $\log R$ and $\log I$ and present our new Bayesian interpretation of $\log R$.
In \cref{sec:analytic} we examine analytic examples to aid intuition on the properties of the Bayes and information ratios.
In \cref{sec:numeric} we apply our techniques to cosmological datasets, with our key results reported in \cref{tab:results}. We conclude in \cref{sec:conclusion}.

\begin{figure}
    \includegraphics{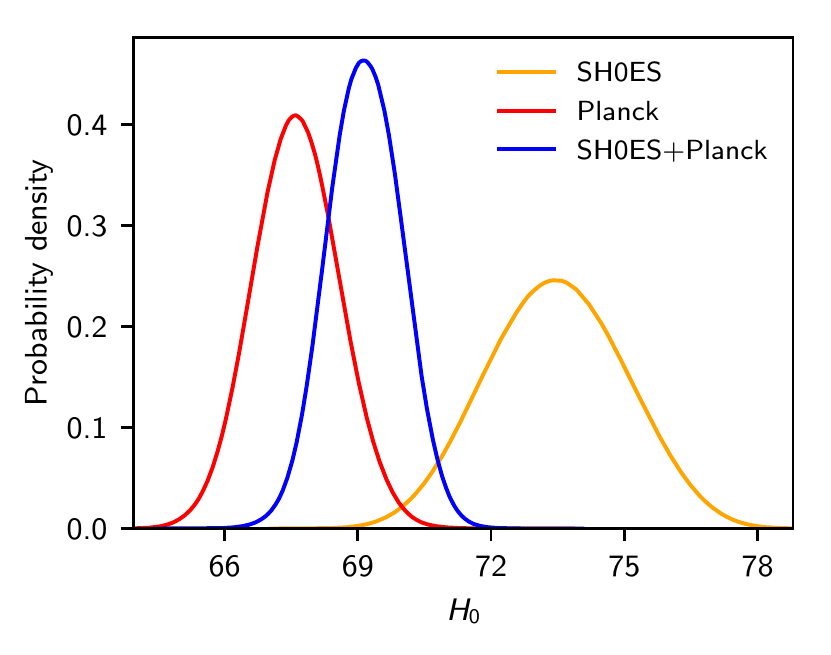}
    \caption{Tension between the \SHOES{} and \Planck{} datasets as exhibited by examining the posterior parameter constraints on the Hubble constant.\label{fig:H0}}
\end{figure}

\section{Background}\label{sec:background}
In general we use the following notation for the quantities in Bayes' theorem:
\begin{equation}
    P(\theta|D) = \frac{P(D|\theta)P(\theta)}{P(D)} \quad\Leftrightarrow\quad
    \mathcal{P}_D(\theta) = \frac{\mathcal{L}_D(\theta)\pi(\theta)}{\mathcal{Z}_D},
    \nonumber
\end{equation}
namely, the posterior $\mathcal{P}$, likelihood $\mathcal{L}$, prior $\pi$, and evidence $\mathcal{Z}$. We will retain dataset-dependence as a subscript, and in general will suppress explicit dependency on $\theta$ except where its presence increases clarity. Furthermore there is a suppressed explicit model dependence, which is taken to be $\Lambda$CDM for our cosmological examples.

\subsection{Bayesian evidence}
Throughout this paper the Bayesian evidence $\mathcal{Z}$, defined as
\begin{equation}
    \mathcal{Z}_D = \int \mathcal{L}_D\pi \d{\theta},
\end{equation}
will play a key role. Also known as the marginal likelihood~\cite{Trotta2008}, the evidence is a key element of model comparison, and may be computed analytically in some rare cases, but is usually evaluated using a Laplace approximation~\cite{Mackay2002}, Savage Dickey ratio~\cite{SavageDickey}, or better still with numerical evidence calculators such as \texttt{MCEvidence}~\cite{MCEvidence1,MCEvidence2} or nested sampling~\cite{Skilling2006,Feroz2008,PolyChord0,PolyChord1,DNest,DNest4}.

\subsection{Kullback-Leibler divergence}
The Kullback-Leibler divergence~\cite{Kullback:1951} is defined as
\begin{equation}
    \mathcal{D}_D = \int \mathcal{P}_D(\theta) \log \frac{\mathcal{P}_D(\theta)}{\pi(\theta)} d{\theta} = \left\langle\log \frac{\mathcal{P}_D}{\pi}\right\rangle_{\mathcal{P}_D,}
    \label{eqn:KL}
\end{equation}
which quantifies the information gain/compression between prior and posterior and has been used by numerous authors~\cite{Hoyosa:2004,Verde:2013,Seehars2014,Seehars2016,Grandis2016,Raveri2016b,HthreeL,Grandis2016b,Zhao2017,Nicola2017,Nicola2019}. The angular brackets $\langle f\rangle_p$ in the right-most expression of \cref{eqn:KL} denote the average of $f$ over the distribution $p$.

\subsection{Bayesian model dimensionality}
We define the Bayesian model dimensionality~\cite{ModelDim} as
\begin{equation}
    \frac{\tilde{d}_D}{2} = \left\langle{\left(\log\frac{\mathcal{P}_D}{\pi}\right)}^2\right\rangle_{\mathcal{P}_D}-\quad {\left\langle\log\frac{\mathcal{P}_D}{\pi}\right\rangle}_{\mathcal{P}_D.}^2
    \label{eqn:d}
\end{equation}
The quantity $\log[\mathcal{P}_D(\theta) / \pi(\theta)]$ is the Shannon information~\cite{Shannon:1949} provided by the posterior relative to the prior at parameter $\theta$, measured in nats (natural bits). As can be seen from \cref{eqn:KL}, the Kullback-Leibler divergence is the average amount of information provided by the posterior, whilst \cref{eqn:d} shows that the Bayesian model dimensionality is proportional to the variance of the information provided by the posterior. 

It should be noted that an earlier preprint of this paper used an alternative definition of the dimensionality by Spiegelhalter~\cite{Spiegelhalter}, which has several unattractive theoretical qualities when applied to significantly non-Gaussian cases. The fundamental qualitative conclusions remain unchanged from the initial version of this paper, and the newer definition of model dimensionality is examined in greater detail in~\cite{ModelDim}.

\begin{figure}
    \includegraphics{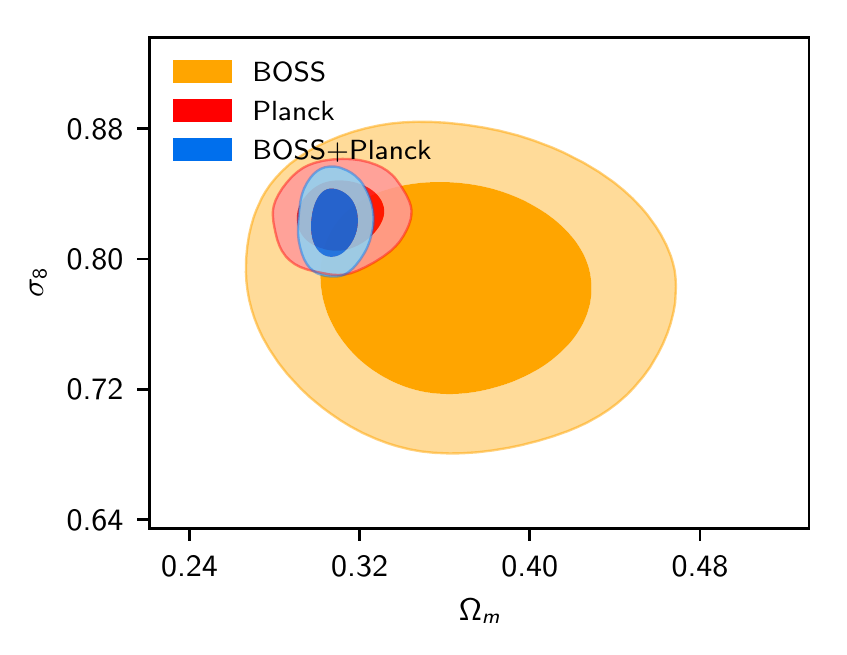}
    \caption{No tension between \SDSS{} and \Planck{} datasets as exhibited by examining the joint posterior parameter constraints on the matter fraction and $\sigma_8$. \label{fig:bao}}
\end{figure}
\subsection{Combining likelihoods}
Independent datasets $A$ and $B$ are combined at the likelihood level via $\mathcal{L}_{AB} =\mathcal{L}_{A}\mathcal{L}_{B}$ so that
\begin{gather}
    \mathcal{P}_A = \frac{\mathcal{L}_A\pi}{\mathcal{Z}_A},
    \quad
    \mathcal{P}_B = \frac{\mathcal{L}_B\pi}{\mathcal{Z}_B},
    \quad
    \mathcal{P}_{AB} = \frac{\mathcal{L}_A\mathcal{L}_B\pi}{\mathcal{Z}_{AB}}.
    \label{eqn:Pdef}
    \\
    \mathcal{Z}_A = \int\mathcal{L}_A\pi\d{\theta},
    \quad
    \mathcal{Z}_B = \int\mathcal{L}_B\pi\d{\theta},\nonumber\\
    \mathcal{Z}_{AB} = \int\mathcal{L}_A\mathcal{L}_B\pi\d{\theta}.
    \label{eqn:Zdef}
\end{gather}
In general, new datasets may introduce additional parameters, either because more cosmological parameters are constrained, or because additional nuisance parameters associated with foregrounds or instrumentation are required to perform inference. In general $\theta$ will be taken to be the span of the entire parameter space of interest.

An important point, often misunderstood by professional practitioners, is that the introduction of unconstrained parameters should not impact on proper inference. It is oft-quoted that Bayes factors (or equivalently evidences) penalise additional parameters, but in fact Bayes factors only penalise constrained parameters. For example, if one were to perform a model comparison between the six-parameter $\Lambda$CDM model and an extension to the model which factored in the age of the cosmologist doing the calculation, then both models would have the same evidence value, since a cosmologist's age is (almost) completely unconstrained by cosmological likelihoods. This is not a bug, but a desirable feature of Bayes factors in their use in consistent inference. The proper Bayesian way to deal with this apparent problem is to exclude such trite models at the model prior level.

\section{The $R$ statistic}\label{sec:R}

\subsection{Definition and prior-dependence}

Given two datasets $A$ and $B$, the $R$ statistic is defined via the equivalent expressions:
\begin{equation}
    R = \frac{\mathcal{Z}_{AB}}{\mathcal{Z}_A \mathcal{Z}_B} = \frac{P(A,B)}{P(A)P(B)} = \frac{P(A|B)}{P(A)} = \frac{P(B|A)}{P(B)},
    \label{eqn:R}
\end{equation}
with all probabilities implicitly conditional on an underlying model (e.g. $\Lambda$CDM). A value of $R \gg 1$ is interpreted as both datasets being consistent, while $R \ll 1$ means the datasets are inconsistent. Note that whilst we assume that the datasets $A$ and $B$ are independent, this does not imply that $R=1$. Specifically, dataset independence means that likelihoods $\mathcal{L}_D(\theta)=P(D|\theta)$, which are probabilities conditioned on $\theta$, combine by multiplication, but evidences $\mathcal{Z}_D = P(D)$, which are likelihoods marginalised over the prior $\pi(\theta)=P(\theta)$, do not.

In the DES Y1 analysis, $R$ is used to quantify tension, with the Jeffreys' scale used as the arbiter for whether models are consistent or not. The interpretation on a Jeffreys scale is somewhat unjustified, as the \DES{} papers do not explain which probability ratio they are placing on the scale.

A second, arguably larger concern is that whilst $R$ satisfies many of desiderata that one would hope for from such a quantity (dimensional consistency, symmetry, parameterisation invariance, use of Bayesian quantities), it is strongly prior-dependent.
We can render this dependency explicit by combining~\cref{eqn:R,eqn:Zdef,eqn:Pdef} to yield:
\begin{equation}
    R=\int \frac{\mathcal{P}_A \mathcal{P}_B}{\pi} \d{\theta} = \left\langle\frac{\mathcal{P}_B}{\pi}\right\rangle_{\mathcal{P}_A}= \left\langle\frac{\mathcal{P}_A}{\pi}\right\rangle_{\mathcal{P}_B}.
    \label{eqn:identities}
\end{equation}
Thus, $R$ can be thought of as the posterior average of the ratio of the other posterior to the shared prior.
More specifically, $R$ depends on the priors set on constrained parameters shared between likelihoods, but not on the prior on additional nuisance or unconstrained parameters. 

It should be noted that this variation is in opposition to the usual evidence prior-dependency. Namely, reducing the widths of the prior in general increases evidence. The same reduction of prior widths however will {\em reduce\/} the ratio $R$ and increase tension. This is easily understood, since in the $R$ ratio there are two evidences on the denominator with only one in the numerator. In a Bayesian sense this is an attractive balance---you can only evidence-hack at the expense of tension. 

It is important to note that the prior dependence of $R$ can only hide existent discordance, i.e. $R$ can indicate that two datasets are in agreement, even when they are not. However, if $R$ indicates that two datasets are discordant, this should be taken seriously, since the prior volume effect only increases the value of $R$.

\begin{figure}
    \includegraphics{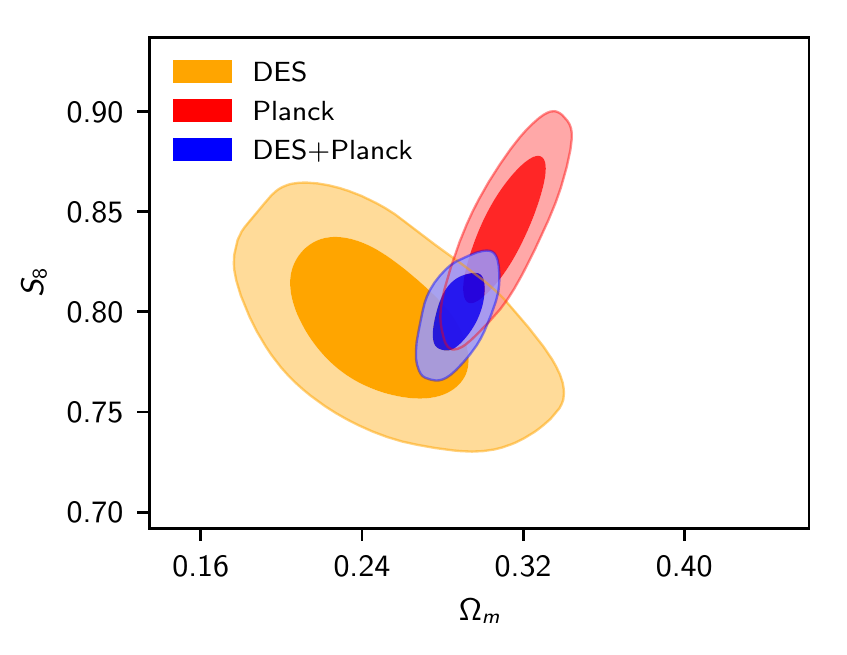}
    \caption{Possible tension between \DES{} and \Planck{} datasets as exhibited by examining the joint posterior parameter constraints on the matter fraction and the parameter combination $S_8 = \sigma_8 (\Omega_m/0.3)^{0.5}$. \label{fig:s8}}
\end{figure}
\subsection{Bayesian interpretation of $R$}

An interpretation that is often posited is that $R$ represents a ratio of probabilities that the shared model parameters come from different universes in comparison with the probability that they come from the same universe. Given that evidences are traditionally used in the context of model comparison, this seems a natural interpretation. However, in order to convert evidences to model probabilities, one requires model priors and for probabilities to be conditioned on the same dataset, which in this case is not true. Raw evidences are probabilities of {\em data}, not of models.

A correct interpretation can be found by examining the two right-hand most expressions in~\cref{eqn:R}. These expressions show that $R$ represents the relative confidence that we have in dataset $A$ in light of knowing dataset $B$, compared to the confidence in $A$ alone (and vice-versa). If $R>1$, then $B$ has strengthened our confidence in $A$ by a factor $R$. If $R\ll1$, then as Bayesians we should be concerned that there is either a problem with the underlying model, or a problem with either or both of the datasets, and therefore avoid combining the two.

Given this interpretation, it is important to understand the prior-dependency of $R$, namely that decreasing the prior widths on shared parameters reduces our confidence in the ability to combine datasets. 

If a Bayesian specifies extremely wide and uniform priors, they are saying that they {\em a priori\/} believe the parameter constraints derived from a dataset $D$ could reside anywhere within that region. It is therefore reassuring when two independent datasets result in constraints that are close. We should be proportionally more reassured if our initial prior were wider, as it is proportionally less likely {\em a priori\/} that they should lie close to one another.

Some practitioners might consider this prior dependency pathological, rather than the correct behaviour of such a probability. In our experience, the primary difference between full Bayesians and other statisticians is that a Bayesian considers this kind of prior-dependent behaviour of the analysis a feature rather than a bug.

Given this prior dependency and its sensible interpretation, the approach we advocate is as follows:
\begin{proposal}\label{proposal1}
    If there are {\em any\/} physically reasonable priors which render $R$ significantly less than 1, then as Bayesians we should consider these datasets in tension. 
\end{proposal}

Given that narrowing the priors decreases the value of $R$, the physically reasonable priors that render the lowest possible value of $R$ are the narrowest priors that do not significantly alter the shape of the posteriors. Whilst such an extreme strategy would provide a definitive lower bound on $R$, many Bayesians would disagree with such a procedure, as it uses a prior that depends on the posterior. In reality, the most pragmatic approach is to choose reasonable initial priors, and then to examine the sensitivity of the conclusions to reasonable alterations to them.

\subsection{Information and suspiciousness}
The logarithmic version of  \cref{eqn:R} for the Bayes ratio in between two datasets $A$ and $B$ is defined as
\begin{equation}
    \log R = \log\mathcal{Z}_{AB} - \log\mathcal{Z}_A - \log\mathcal{Z}_B.
    \label{eqn:logR}
\end{equation}
As discussed in the previous section, the Bayesian confidence $R$ has two primary contributions, one from the unlikeliness of two datasets ever matching (proportional to prior), and another in their mismatch.
We may quantify the first of these via the {\em information ratio\/} $I$ defined using Kullback-Leibler divergences as:
\begin{equation}
    \log I = \mathcal{D}_A+\mathcal{D}_{B} - \mathcal{D}_{AB}.
    \label{eqn:logI}
\end{equation}
The remaining part of the Bayesian confidence quantifies the mismatch, which we term the {\em suspiciousness} $S$:
\begin{equation}
    \log S = \log R - \log I.
    \label{eqn:logS}
\end{equation}
Suspiciousness is unaffected by changing the prior widths as long as this change does not significantly alter the posterior, since the information ratio $I$ and Bayes ratio $R$ transform similarly under prior volume alterations. 

It is important to recognise that whilst $\log S$ is indeed prior-independent, in constructing this quantity we have lost the probabilistic interpretation found in $\log R$. More care must be taken to calibrate the scale on which $\log S$ sits, which will be considered at the end of the next section.

\section{Analytical examples}\label{sec:analytic}

In all of the below, for a graphical understanding, one may substitute $A$ $\leftrightarrow$ \Planck{}, $B$ $\leftrightarrow$ \SHOES{}, \DES{}, or \SDSS{}  and consult \cref{fig:s8,fig:H0,fig:bao} respectively.

For simplicity, we consider $A$ and $B$ to have the same parameters $\theta$, although the case is easily extended to the case where the likelihoods only share some parameters, in which case our results depend only on those parameters that are shared between likelihoods.

\subsection{Top-hat example}

As a simple choice, we consider a top-hat posterior over a multidimensional region $\mathcal{R}_X$, enclosing a volume $V_X$:
\begin{equation}
    \mathcal{P}_X(\theta) = 
    \left\{
    \begin{array}{cc}
        V_X^{-1} &: \theta\in \mathcal{R}_X \\
        0 &: \text{otherwise}
    \end{array}
    \right.
    ,
    \qquad
    V_X = \int_{\theta\in \mathcal{R}_X}\d{\theta}.
    \label{eqn:uniform}
\end{equation}
If we have a top-hat prior with volume $V_\pi$ enclosing two top-hat posteriors $P_A$ and $P_B$, along with their combined posterior $P_{AB}$, then 
\begin{align}
    \log R = \log I = \log\frac{V_{AB}V_\pi}{V_A V_B}.
\end{align}
We can see the explicit prior dependency of $R$ with the presence of the $V_\pi$ term. Furthermore, we see that $R$ and $I$ are equal in the top-hat posterior, so that the entire contribution to $R$ is in information, and none in suspicion:
\begin{align}
    S = 
    \left\{
    \begin{array}{cc}
        1 &: \mathcal{R}_A \cap \mathcal{R}_B \ne \emptyset \\
        0 &: \text{otherwise.}
    \end{array}
    \right.
    \label{eqn:logS_uniform}
\end{align}
Thus for the uniform case there is no suspiciousness, provided that the posteriors have any overlap region and are thus plausibly consistent.


\subsection{Gaussian example}
\begin{figure*}
    \includegraphics[width=0.49\textwidth]{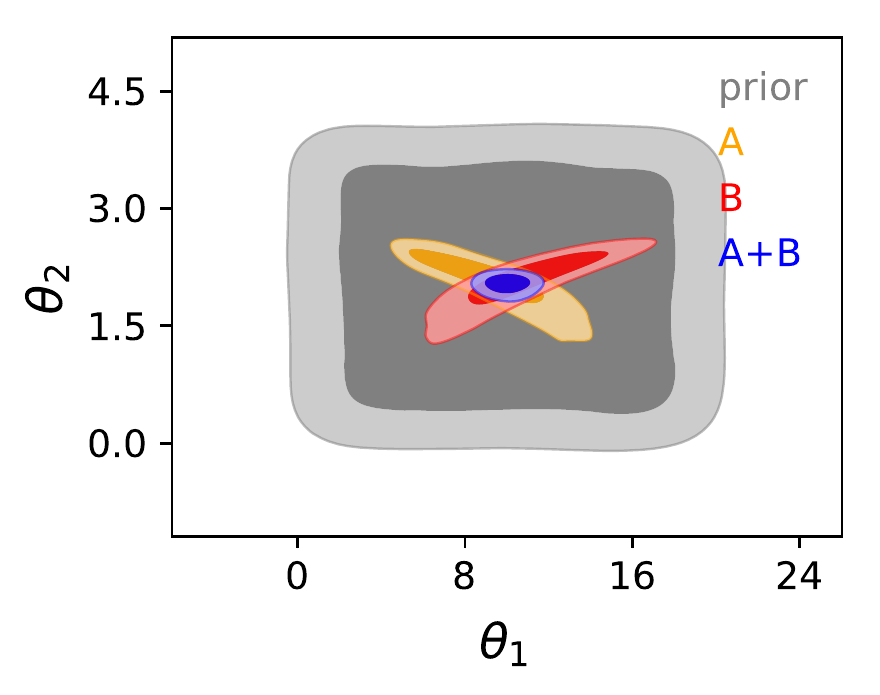}
    \includegraphics[width=0.49\textwidth]{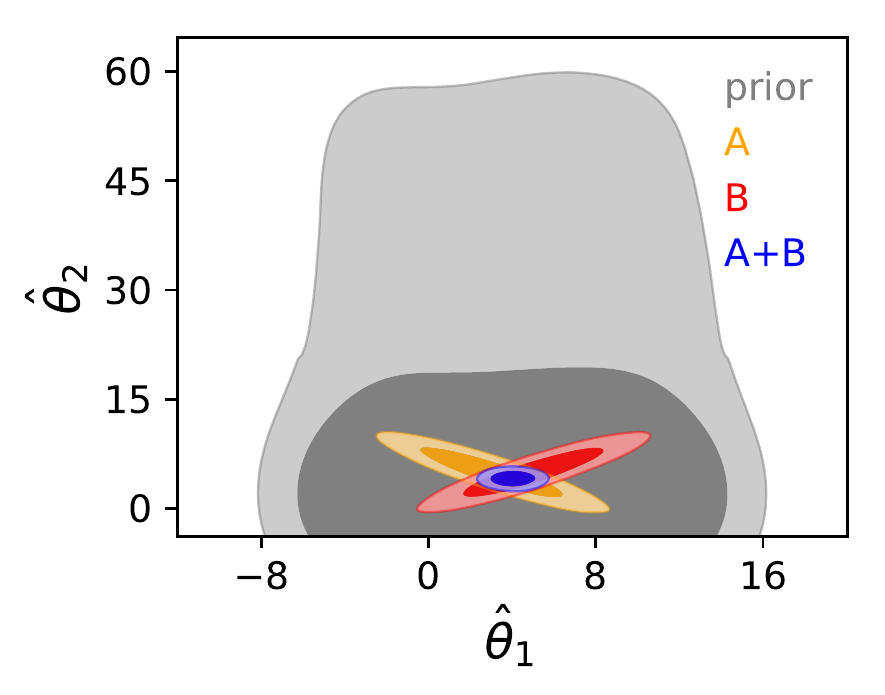}
    \caption{Many non-Gaussian posteriors (left) may be ``Gaussianised'' (right) by using Box-Cox transformations.\label{fig:box_cox}}
\end{figure*}

We now consider a less trivial multivariate Gaussian example~\cite{Hubblehierarchy,Feeney2018}. A $d$-dimensional Gaussian likelihood with peak $\mathcal{L}_\mathrm{max}$, centre $\mu$ and parameter covariance $\Sigma$, along with a top-hat enclosing prior over volume $V_\pi$ has likelihood, posterior, evidence and Kullback-Leibler divergence given by the following:
\begin{align}
    \log\mathcal{L}(\theta) &= \log\mathcal{L}^\mathrm{max} - \frac{1}{2}(\theta-\mu) \Sigma^{-1}(\theta-\mu),
\label{like1}\\
\log\mathcal{P}(\theta) &= -\frac{1}{2}\log |2\pi\Sigma| - \frac{1}{2}(\theta-\mu) \Sigma^{-1}(\theta-\mu),
    \\
    \log \mathcal{Z} &= \log\mathcal{L}^\mathrm{max} + \frac{1}{2}\log |2\pi\Sigma| - \log V_\pi,
    \\
    \mathcal{D} &=   \log V_\pi - \frac{1}{2}(d+\log |2\pi\Sigma|).
\end{align}
Note that in the above we have removed explicit dimensionality-dependency from the normalisation of a Gaussian by exploiting the matrix determinant property $|2\pi\Sigma| = {(2\pi)}^{d}|\Sigma|$.

Two likelihoods $A$ and $B$ combine using the relations
\begin{align}
    \log\mathcal{L}^\mathrm{max}_{AB} =& -\frac{1}{2} (\mu_A-\mu_B){(\Sigma_{A}+\Sigma_{B})}^{-1}(\mu_A-\mu_B) \nonumber\\
    & +\log\mathcal{L}^\mathrm{max}_{A} + \log\mathcal{L}^\mathrm{max}_{B}, \\
    \Sigma_{AB}^{-1} =& \Sigma_A^{-1} + \Sigma_B^{-1},\\
    \mu_{AB} =& \Sigma_{AB}\left[ \Sigma_A^{-1}\mu_A + \Sigma_B^{-1}\mu_B \right].
\end{align}
It should also be noted that
\begin{equation}
    {(\Sigma_A+\Sigma_B)}^{-1} = \Sigma_A^{-1} \Sigma_{AB} \Sigma_B^{-1} = \Sigma_B^{-1} \Sigma_{AB} \Sigma_A^{-1}.
\end{equation}
We therefore find
\begin{align}
    \log R =& -\frac{1}{2} (\mu_A-\mu_B){(\Sigma_{A}+\Sigma_{B})}^{-1}(\mu_A-\mu_B)
    \nonumber\\
    &  -\frac{1}{2}\log\left|2\pi(\Sigma_{A}+\Sigma_{B})\right|  + \log V_\pi,
\end{align}
and 
\begin{align}
    \log I =&  - \frac{d}{2} -\frac{1}{2}\log\left|2\pi(\Sigma_{A}+\Sigma_{B})\right|  + \log V_\pi.
\end{align}
We thus find the information content can be used to remove all of the residual prior dependence from $\log R$, giving a suspiciousness:
\begin{equation}
    \log S = \frac{d}{2}  -\frac{1}{2} (\mu_A-\mu_B){(\Sigma_{A}+\Sigma_{B})}^{-1}(\mu_A-\mu_B).
\end{equation}
The numerical value of the suspiciousness is determined by the means and covariances of the posterior distributions $A$ and $B$. Under a Bayesian interpretation of the posterior, if the ``true'' value of the measured parameter is $\theta_0$, then both means are drawn from a normal distribution centred on this value with covariance equal to their posterior covariance $\mu_A \sim \mathcal{N}(\theta_0, \Sigma_A)$, $\mu_B \sim \mathcal{N}(\theta_0, \Sigma_B)$, and their difference is drawn from a distribution centred on zero with covariance equal to the sum of the underlying covariances $\mu_A - \mu_B \sim \mathcal{N}(0, \Sigma_A + \Sigma_B)$. One can see that $d-2\log S$, has a $\chi^2_d$ distribution, and that $\log S$ is typically $0\pm \sqrt{d/2}$. An overly negative value of $\log S$ indicates discordance, and an overly positive value suspicious concordance. More quantitatively, one can use the inverse cumulative $\chi^2_d$ distribution to turn $\log S$ into the {\em tension probability\/} of two datasets being this discordant by chance:
\begin{equation}
    p = \int\limits_{d-2\log S}^\infty \chi^2_d(x) \d{x} = \int\limits_{d-2\log S}^\infty \frac{x^{d/2-1}e^{-x/2}}{2^{d/2}\Gamma(d/2)} \d{x}.
    \label{eqn:p}
\end{equation}

Whilst this procedure is only exact for the Gaussian case, a reasonable proposition for general posteriors would be to compute $\log S$ numerically, and then determine tension via a $\chi^2$-like test, in analogy with the Gaussian case:
\begin{proposal}\label{proposal2}
    If $p\lesssim0.05$, where $p$ is the tension probability computed from \cref{eqn:p}, $\log S$ is computed using numerical evidences and Kullback-Leibler divergences, and $d=\tilde{d}_A+\tilde{d}_B-\tilde{d}_{AB}$ is the Bayesian model dimensionality of the shared constrained parameters computed using \cref{eqn:d}, then the datasets should be considered in moderate tension. If $p \lesssim 0.003$, they should be considered in strong tension.\footnote{$p=0.05$ and $0.003$ correspond to $2$- and $3$-$\sigma$ Gaussian standard deviations.}
\end{proposal}

For the case when the posteriors are exactly (or extremely close to) Gaussian, the tension probability $p$ may be interpreted as a probability that one would observe such a discrepancy by chance alone. In the non-Gaussian case, $p$ is only a rough calibration so only extremely small values of $p$ should be regarded with suspicion. The suspiciousness $S$ can be used to determine discordance if $S\ll -\sqrt{\tilde{d}/2}$, and the tension probability $p$ provides a mechanism for putting a number on the concept of $\ll$ in this case. The $R$ statistic, however, is always interpretable as a Bayesian confidence in our ability to combine the data, irrespective of Gaussianity. 

It should be noted that many posteriors may be ``Gaussianised'' using techniques like Box-Cox transformations~\cite{BoxCox}. These transformations are non-linear mappings that can transform complex posteriors into approximately Gaussian ones by changing the parameterization, and have already been used in the context of cosmology~\cite{Joachimi2011, hiranya_box_cox}. It can be easily proven that these transformations preserve the value of the suspiciousness, although care must be taken to also transform the underlying common prior distribution appropriately (\cref{fig:box_cox}), and that the prior is not significantly distorted by the Box-Cox transformation in the region of the posterior bulk.

Our two propositions for tension quantification are in fact related: one can think of $\log I$ as being the volume of the narrowest prior that does not significantly impinge upon the posterior bulk, and \cref{proposal2} is one method for quantifying the qualitative statement ``any reasonable prior'' in \cref{proposal1}. Finally, the interpretation of the Bayesian model dimensionality $\tilde{d}_D$ as the effective number of parameters is made clear in the Gaussian case, since $\tilde{d}_D=d$.

\section{Numerical examples}\label{sec:numeric}

\begin{table*}

    {
        \setlength{\tabcolsep}{0.6em}
    
    \begin{tabular}{|c|ccccccc|}
\hline
        Prior & \SHOES{} &        \SDSS{} &         \DES{} &          \Planck{} &       \SHOES{}+\Planck{} & \SDSS{}+\Planck{} & \DES{}+\Planck{} \\
\hline
 default\cellcolor{red!20}   & $    0.93 \pm     0.03$ & $    2.95 \pm     0.07$ & $   14.01 \pm     0.32$ & $   15.84 \pm     0.38$ & $   15.98 \pm     0.37$ & $   15.89 \pm     0.36$ & $   25.89 \pm     0.63$\\ 
\hline
 medium\cellcolor{orange!20} & $    0.98 \pm     0.03$ & $    3.79 \pm     0.09$ & $   13.35 \pm     0.31$ & $   15.89 \pm     0.38$ & $   15.09 \pm     0.35$ & $   16.38 \pm     0.37$ & $   26.10 \pm     0.65$\\ 
\hline
 narrow\cellcolor{blue!20}   & $    1.68 \pm     0.03$ & $    1.40 \pm     0.02$ & $   10.89 \pm     0.24$ & $   15.96 \pm     0.37$ & $   15.72 \pm     0.37$ & $   15.69 \pm     0.37$ & $   25.69 \pm     0.62$\\ 
\hline
    \end{tabular}
}
 
\caption{Bayesian model dimensionality of $\Lambda$CDM for all datasets and priors considered in this paper, calculated using \cref{eqn:d}.\label{tab:dim}}
\end{table*}

\begin{table*}
    {
        \setlength{\tabcolsep}{0.6em}
    \begin{tabular}{|c|c|rrrrr|}

        \hline
        Dataset            & Prior & $\hfill\log R\hfill$ & $\hfill\log I\hfill$ & $\hfill\log S\hfill$ & $\hfill \tilde{d}\hfill$ & $\hfill p (\%)\hfill$ \\
        \hline
        \SDSS{}-\Planck{}  & default\cellcolor{red!20}    & $    6.30 \pm     0.29$ & $    6.18 \pm     0.29$ & $    0.11 \pm    0.11$ & $    2.91 \pm     0.51$ & $   42.66 \pm     4.28$ \\ 
                           & medium\cellcolor{orange!20}  & $    4.51 \pm     0.28$ & $    4.06 \pm     0.28$ & $    0.46 \pm    0.12$ & $    3.30 \pm     0.55$ & $   55.12 \pm     4.47$ \\ 
                           & narrow\cellcolor{blue!20}    & $    1.30 \pm     0.23$ & $    0.69 \pm     0.22$ & $    0.61 \pm    0.12$ & $    1.67 \pm     0.54$ & $   77.12 \pm    14.10$ \\ 
        \hline                                                                                                                                                                                
        \DES{}-\Planck{}   & default\cellcolor{red!20}    & $    2.88 \pm     0.35$ & $    6.15 \pm     0.34$ & $   -3.28 \pm    0.16$ & $    3.97 \pm     0.82$ & $    3.23 \pm     1.00$ \\ 
                           & medium\cellcolor{orange!20}  & $    0.51 \pm     0.34$ & $    4.00 \pm     0.34$ & $   -3.49 \pm    0.16$ & $    3.13 \pm     0.81$ & $    2.04 \pm     0.79$ \\ 
                           & narrow\cellcolor{blue!20}    & $   -1.88 \pm     0.29$ & $    0.90 \pm     0.29$ & $   -2.78 \pm    0.16$ & $    1.15 \pm     0.77$ & $    1.44 \pm     0.91$ \\ 
        \hline                                                                                                                                                                                
        \SHOES{}-\Planck{} & default\cellcolor{red!20}    & $   -2.03 \pm     0.29$ & $    1.96 \pm     0.28$ & $   -3.99 \pm    0.12$ & $    0.78 \pm     0.52$ & $    0.25 \pm     0.17$ \\ 
                           & medium\cellcolor{orange!20}  & $   -2.50 \pm     0.28$ & $    1.56 \pm     0.28$ & $   -4.06 \pm    0.11$ & $    1.77 \pm     0.51$ & $    0.56 \pm     0.24$ \\ 
                           & narrow\cellcolor{blue!20}    & $   -2.00 \pm     0.23$ & $    1.43 \pm     0.23$ & $   -3.43 \pm    0.12$ & $    1.92 \pm     0.52$ & $    1.17 \pm     0.45$ \\ 
        \hline
    \end{tabular}
}

\caption{Comparison statistics. The values of $\log R$ and $\log I$ are computed via \cref{eqn:logR,eqn:logI}, using the evidences and Kullback-Leibler divergences reported in \cref{fig:raw}. The suspiciousness statistic is simply $\log S=\log R - \log I$. $\tilde{d}$ is the Bayesian combined model dimensionality from \cref{eqn:d}, detailing the number of shared constrained parameters between the datasets, and $p$ is the tension probability computed from \cref{eqn:p}.  One can see explicitly the prior dependency of $\log R$ and $\log I$, and how this is removed/reduced in $\log S$ and $p$. In both the Bayes ratio $\log R$ via \cref{proposal1} and the tension probability $p$ via \cref{proposal2}, we find that the data show no tension between \SDSS{}-\Planck{}, moderate discordance between \DES{}-\Planck{}, and strong discordance between \SHOES{}-\Planck{}.\label{tab:results}}
\end{table*}

\begin{figure*}
{
    \includegraphics{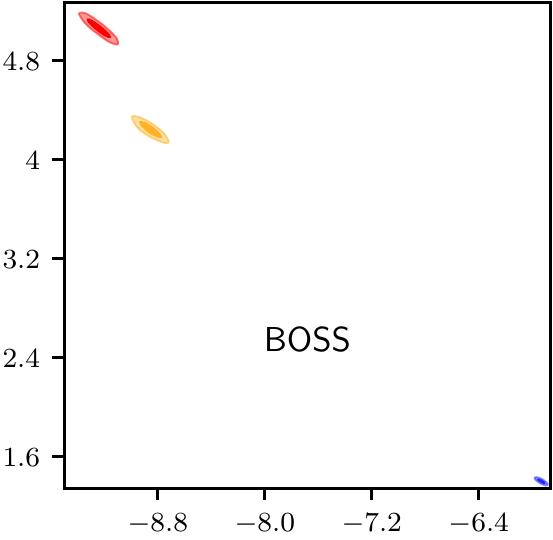}
    \includegraphics{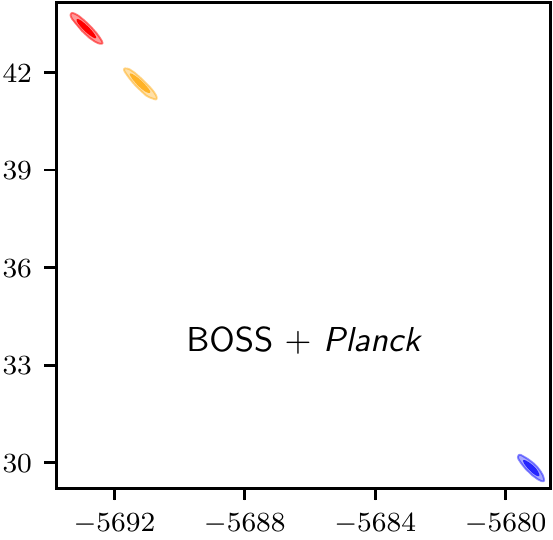}

    \includegraphics{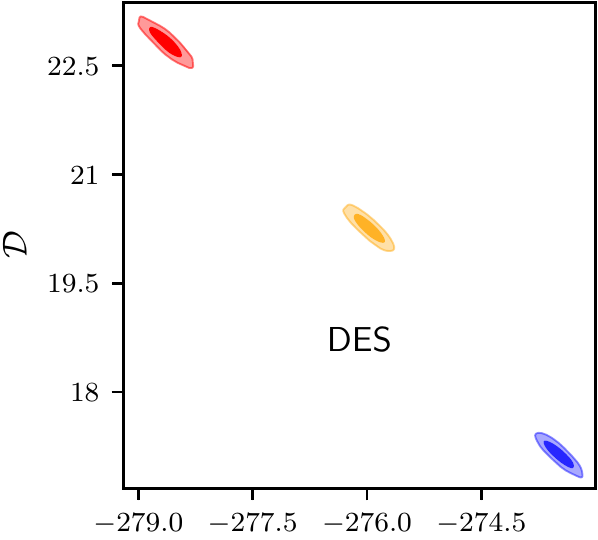}
    \includegraphics{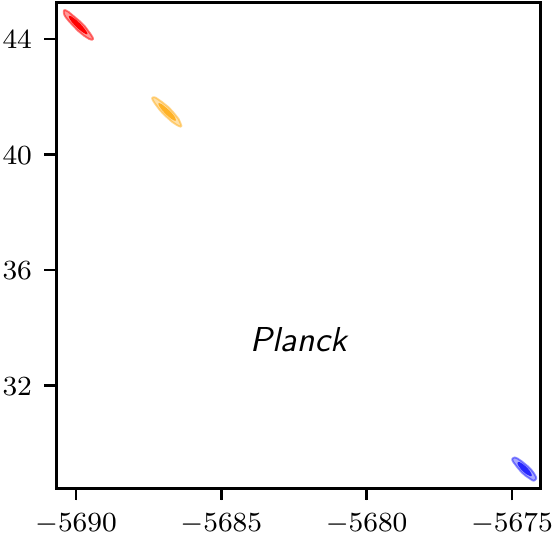}
    \includegraphics{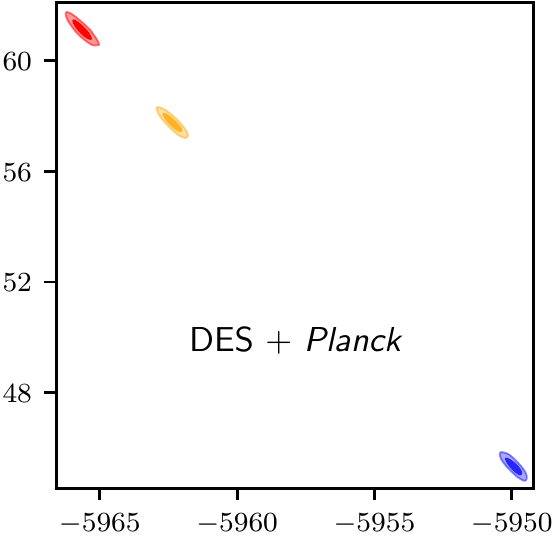}

    \includegraphics{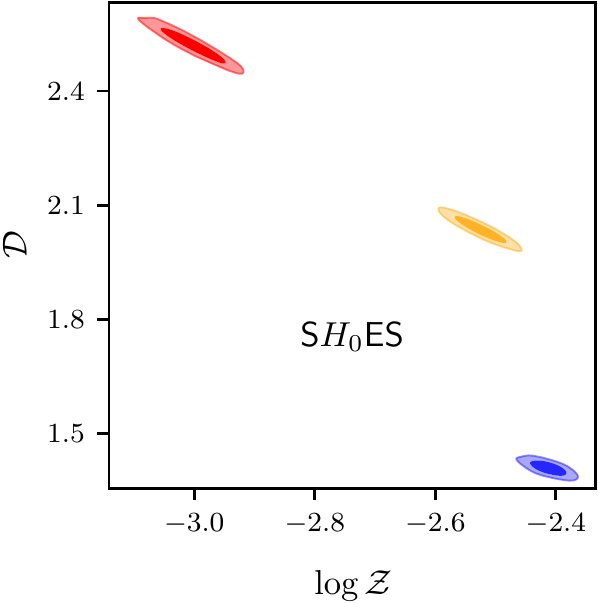}
    \includegraphics{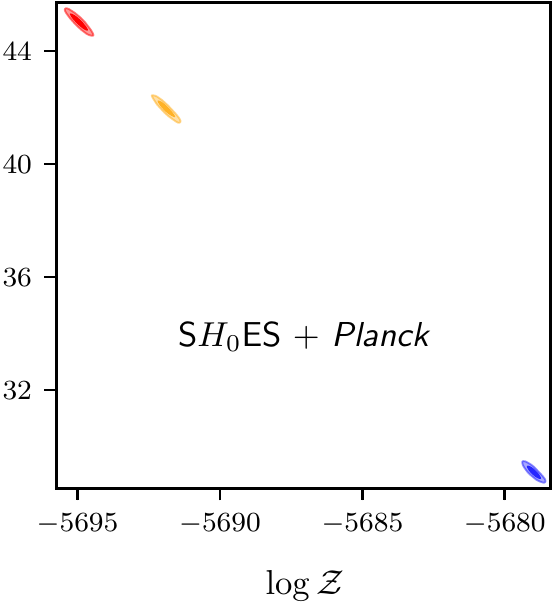}
}
    \caption{Log-evidence $\log \mathcal{Z}$ and Kullback-Leibler divergence $\mathcal{D}$ calculations for all datasets and priors considered in this paper. The figures show the numerical values for the log-evidence and Kullback-Leibler divergence for the likelihoods described in \cref{sec:datasets} under the default and narrow priors summarised in \cref{fig:prior}, with red representing results for the default priors, orange medium priors and blue narrow priors. One can see that narrowing the prior increases the log-evidence and reduces the Kullback-Leibler divergence, but that $\log \mathcal{Z} +\mathcal{D}$ remains constant to within error. It should also be noted that the errors in estimating $\log \mathcal{Z}$ and $\mathcal{D}$ are strongly correlated. These errors arise from the uncertainty inherent in nested sampling's estimate of the volume compression of each likelihood contour, and influences both quantities in the same manner. It should be noted that the parameter combination that we are most interested in estimating ($\log \mathcal{Z} + \mathcal{D}$) has the lowest error in its estimation.\label{fig:raw}}
    \vspace{50pt}
\end{figure*}

We now apply our techniques to the cosmological dataset pairings of Cosmic Microwave Background data (CMB) with Baryon Acoustic Oscillations plus Redshift-Space Distortions (BAO+RSD), galaxy clustering and weak lensing (3x2), and supernovae (SNe) respectively. This necessitates the numerical computation of evidences and Kullback-Leibler divergences via nested sampling. We find that BAO+RSD observations are fully consistent with CMB, 3x2 is in moderate tension, and SNe are in strong tension\@. Our results are summarised in \cref{tab:results}.

\subsection{Nested sampling computation}\label{sec:ns}
To compute the log-evidence $\log \mathcal{Z}$ and the Kullback-Leibler divergence $\mathcal{D}$ we use the outputs of a nested sampling run produced by \CosmoChord{}~\cite{CosmoChord}, a modified version of \CosmoMC{}~\cite{cosmomc} using \PolyChord{}~\cite{PolyChord0,PolyChord1} as a nested sampler. For a reliable computation of evidences and Kullback-Leibler divergences, we found it essential to use \PolyChord{} rather than \MultiNest{}~\cite{Feroz2008}, due to the high dimensionality of the space of cosmological and nuisance parameters\footnote{A little-known test of the reliability of the evidence estimates reported by \MultiNest{} is to check whether two estimates of the evidence (the traditional and importance nested sampling estimation) agree to within the larger error bar. If they do not, then this indicates that the ellipsoidal approximation for generating new live points via rejection sampling is no longer valid. This may be fixed by decreasing the value of the efficiency parameter, with a consequent increase in run time.}. Furthermore, \PolyChord{} is able to dramatically speed up nested sampling in the context of cosmology by utilizing the fast-slow hierarchy between nuisance and cosmological parameters~\cite{cosmomc_fs}. As a historical note, \PolyChord{} was invented as an alternative to \MultiNest{} in the context of the \Planck{} collaboration~\cite{planck_inflation_2015,planck_inflation} to resolve precisely the issues described above.

The log-evidences and KL divergences are computed using the likelihood contours $\mathcal{L}_i$ of the discarded points from the trapezoidal rule 
\begin{align}
    \mathcal{Z} \approx& \sum_{i=1}^{N}\mathcal{L}_i \times\frac{1}{2}(X_{i-1}-X_{i+1}),
    \nonumber\\
    \mathcal{D} \approx& \sum_{i=1}^{N}\frac{\mathcal{L}_i}{\mathcal{Z}}\log\frac{\mathcal{L}_i}{\mathcal{Z}} \times \frac{1}{2} (X_{i-1}-X_{i+1}),\nonumber\\
    \frac{\tilde{d}}{2} \approx& \sum_{i=1}^{N}{\frac{\mathcal{L}_i}{\mathcal{Z}}\left(\log\frac{\mathcal{L}_i}{\mathcal{Z}}-\mathcal{D}\right)}^2 \times \frac{1}{2} (X_{i-1}-X_{i+1}),\nonumber\\
    X_{i} =& t_i X_{i-1}, \qquad X_0 = 1, \qquad X_{N+1}=0,
    \label{eqn:simulate}
\end{align}
where $X_i$ are the prior volumes of the $N$ likelihood contours and the $t_i$ are real random variables with probability distribution function:
\begin{equation}
    P(t_i) = n_{i} t_i^{n_i-1}\:[0<t_i<1]
    \label{eqn:simulate_t}
\end{equation}
Here $n_i$ are the (usually constant) number of active live points enclosed by each likelihood contour $\mathcal{L}_i$. To account for all of the correlation between the random variables $\mathcal{D}$ and $\log \mathcal{Z}$, we simulate a set of weights $\{t_i\}$ using \cref{eqn:simulate_t}, and compute $\mathcal{Z}$, $\mathcal{D}$ and $\tilde{d}$ from \cref{eqn:simulate} using the same weights. This process is repeated 1000 times to build up a set of samples from the $P(\mathcal{Z},\mathcal{D},\tilde{d})$ distribution. Examples of such distributions can be seen graphically in \cref{fig:raw}. The log-sum-exp trick must be carefully utilized to avoid overflow errors throughout these computations. For more detail, consult John Skilling's original nested sampling paper~\cite{Skilling2006}. Code to compute these quantities is now publicly available as part of the \anesthetic{} pip-installable Python package~\cite{anesthetic}.

For our final runs, we used the \CosmoChord{} settings $n_\mathrm{live}=1000$, $n_\mathrm{prior}=10000$, with all other settings left at their defaults for version 1.15. It is worth remarking that run-time is linear in the number of live points, and that \PolyChord{} (in contrast to \MultiNest{}) can function with extremely low numbers of live points. For low-resolution testing purposes $n_\mathrm{live}$ can be set as low as 10, which proves invaluable in the initial exploratory stages of a project when publication-quality runs are not essential.

\subsection{Cosmological Likelihoods}
\label{sec:datasets}
For CMB observations we use the publicly available \Planck{} 2015 TT+lowl+lowTEB likelihoods\footnote{At the time of writing this article, the \Planck{} 2018 likelihoods \cite{PlanckParameters2018} were not publicly available. The main difference between the \Planck{} 2015 and 2018 parameters values is the constraints in the optical depth to reionization $\tau$, that change from $\tau =  0.078 \pm 0.019$ \cite{PlanckParameters2015} to $\tau = 0.055 \pm 0.009$ \cite{PlanckTau2016}. Because this paper is focused on the tension reported in \cite{DESParameters2017}, which uses the \Planck{} 2015 likelihood, including their value of $\tau$, we do not impose any priors on this parameter and simply use the \Planck{} 2015 likelihood. }
~\cite{PlanckLikelihoods2015}. For BAO+RSD observations we use the 6DF+MGS \SDSS{} DR12 final consensus data~\cite{SDSS,SDSS2,SDSS3}. For 3x2 data, we use the 1 year final \DES{} dataset~\cite{DESParameters2017}. Finally, for SNe data we use a Gaussian likelihood on the Hubble parameter with mean and width indicated by the latest \SHOES{} constraints~\cite{Riess2018}.

We follow the notation and parameterisation detailed in the respective likelihood papers, and we direct readers to those for further information on the meaning and notation of parameters.

\subsection{Priors}
\begin{figure*}
    \includegraphics{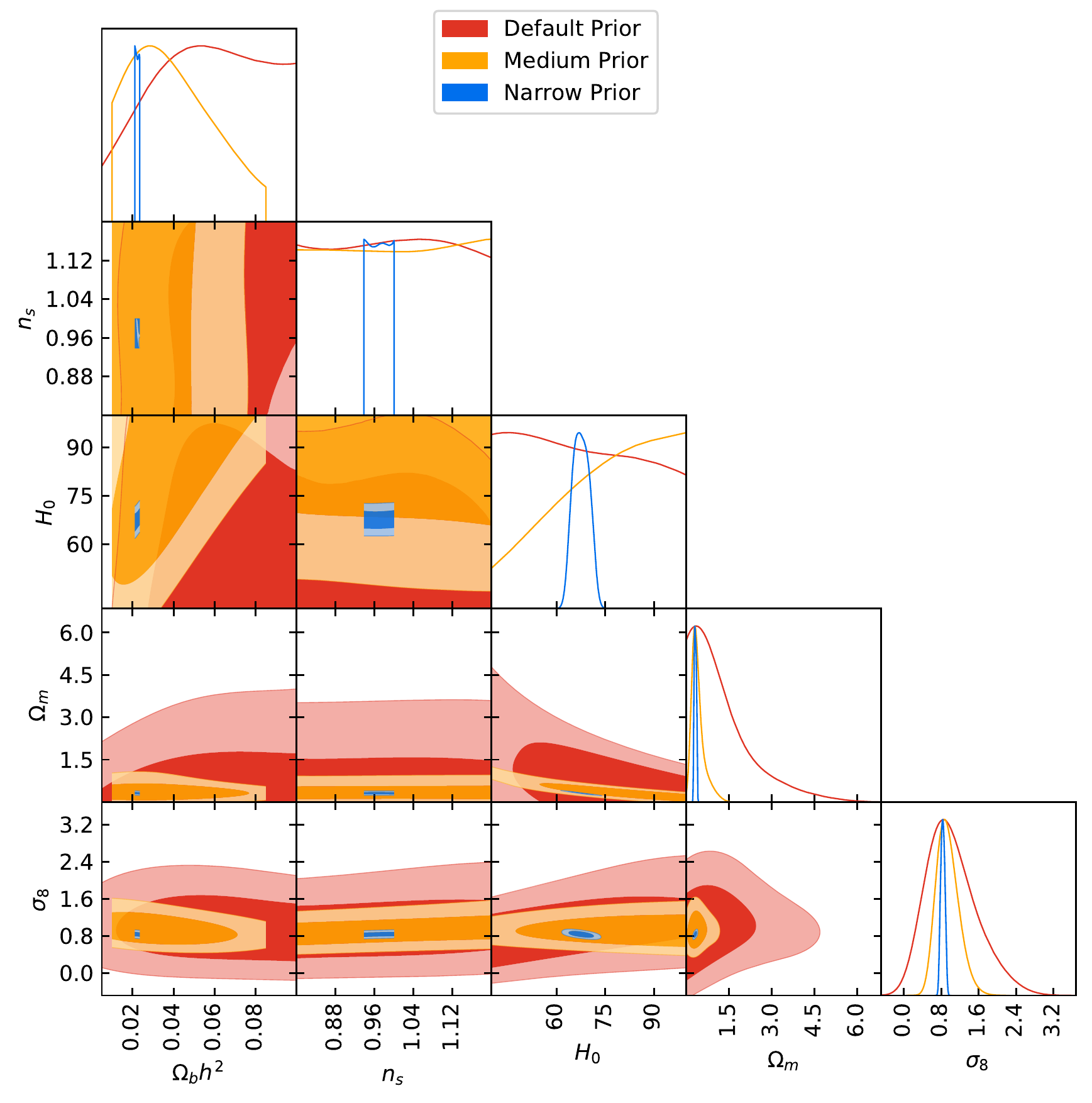}
    \llap{\raisebox{14cm}{
			\begin{tabular}{|llll|}
				\hline
				Parameter                 & Default prior \cellcolor{red!20} & Medium prior \cellcolor{orange!20} & Narrow prior \cellcolor{blue!20} \\
				\hline
				$\Omega_b h^2$            & $[0.005,0.1]$   & $[0.01,0.085]$& $[0.0211,0.0235]$ \\
				$\Omega_c h^2$            & $[0.001,0.99]$  & $[0.08,0.21]$ & $[0.108,0.131]$  \\
				$100\theta_{MC}$          & $[0.5,10]$      & $[0.97,1.5]$  & $[1.038,1.044]$  \\
				$\tau$                    & $[0.01,0.8]$    & $[0.01,0.8]$  & $[0.01,0.16]$  \\
				$n_s$                     & $[0.8,1.2]$     & $[0.8,1.2]$   & $[0.938,1]$  \\
				$\ln(10^{10} A_s)$        & $[1.61,3.91]$   & $[2.6,3.8]$   & $[2.95,3.25]$    \\
				\hline
			\end{tabular}
    }}
    \caption{%
        The three priors used throughout \cref{sec:numeric}. The priors provided to \CosmoMC{} are shown in the upper-right table. These construct an approximate box-prior on the six cosmological parameters. Two of the cosmological parameters are shown in the triangle plot, and indeed a box prior can be seen on the $n_s$ parameter. The $\Omega_b h^2$ prior is not a simple top-hat prior on account of the fact that \CosmoMC{} discards unphysical parameter combinations at the prior level. The remaining parameters are ``derived parameters'' and in general will not have box-priors, as can be seen in the $(H_0,\Omega_b h^2)$ plot. \label{fig:prior}
}
\vspace{50pt}
\end{figure*}

To demonstrate the prior dependencies of $\log R$ and $\log S$, we choose three priors. The first is the default prior provided by \CosmoMC{}. Note that this prior is not a trivial top-hat box prior, since \CosmoMC{} places a model-dependent prior on the parameter space by eliminating regions that are unphysical. This non-trivial shape is shown in \cref{fig:prior}.  
We compare the default with two alternative prior choices; a ``narrow'' box centred on the posterior mean of \Planck{}, with widths extending to 5$\sigma$ of the \Planck{} posterior, and a ``medium'' box designed to encompass the \DES{} posterior whilst being a little narrower than the default. The narrow prior is arguably rather tight, but is chosen as the other extreme end of prior choice from the default prior to emphasise the prior dependency of the $R$ statistic. It is worth noting that there is nothing particularly special about the choice of prior provided by the \CosmoMC{} default, which could easily be narrowed or widened without a great deal of consensus objection.

\subsection{Posteriors}
\begin{figure*}
    \includegraphics[width=0.49\textwidth]{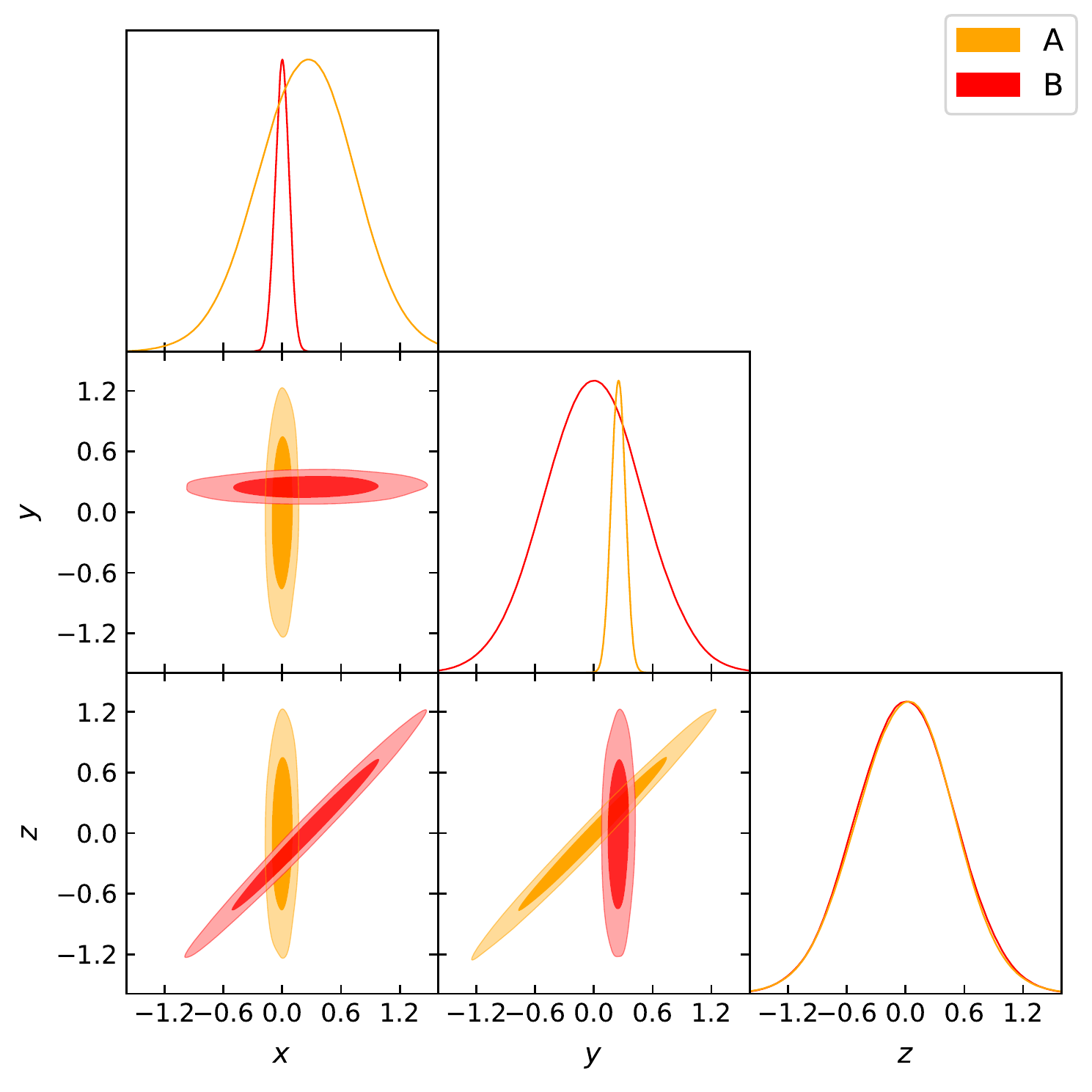}
    \includegraphics[width=0.49\textwidth]{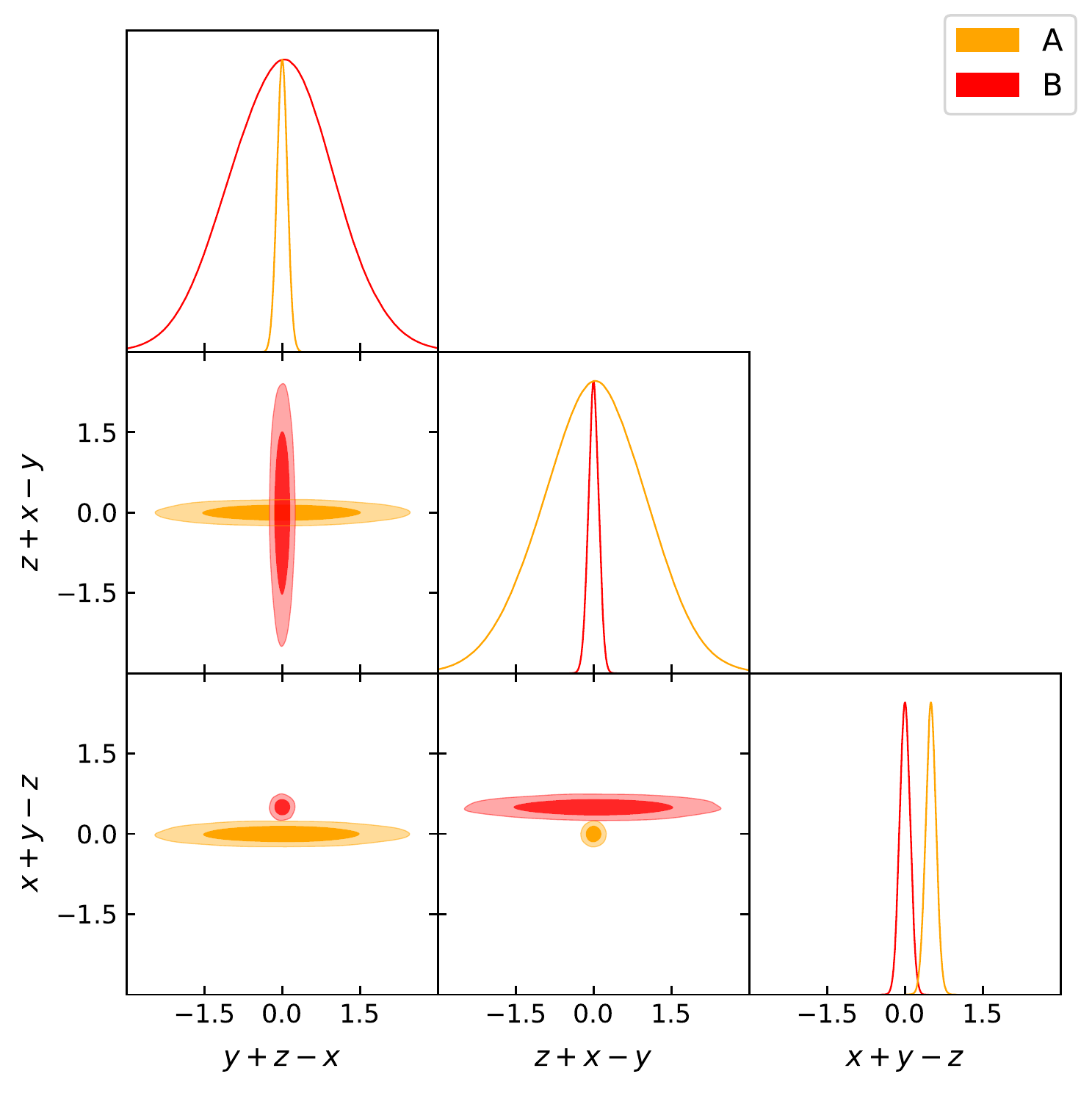}
    \caption{Hidden tension in a multivariate Gaussian. By eye, the three-dimensional posterior on the left seems to be in reasonable agreement. The one- and two-dimensional marginalised posteriors are clearly consistent. However, upon making the linear transformation indicated on the right, it is clear that the posteriors are in fact disjoint. Such issues become harder in higher-dimensional posteriors, and demonstrate the importance of a parameterisation-independent measures of tension such as those that we demonstrate here. \label{fig:rotate}}
\end{figure*}

The posterior on the Hubble parameter for \SHOES{} and \Planck{} produced by \PolyChord{} is shown in \cref{fig:H0}. By eye it is clear from the individual posteriors that the inferences on the value of $H_0$ are incompatible, and that the combined posterior cannot be trusted.

For \SDSS{} and \Planck{}, we show the marginalised posterior on the two parameters $\sigma_8$ and $\Omega_m$ in \cref{fig:bao}. Here there is significant overlap between the two-dimensional marginalised posteriors, and the combined posterior is valid. Note that they do not lie precisely on top on each other, which is in itself reassuring as otherwise the datasets would be suspiciously in agreement (and would usually indicate an overestimate of the errors or biases in the analysis).

For \DES{} and \Planck{}, we show the marginalised posterior for two parameters similar to those used in the \SDSS{} case. In this case the situation is less clear, with a large proportion of the marginalised posterior bulk in disagreement, but with a small degree of overlap. If one looks at other parameter combinations, the tension becomes better or worse, and indeed it is possible to consider situations where there appears to be excellent overlap in every pair of parameters. However, it should be noted that since tension is a parameter invariant notion, if one can resolve a significant tension in {\em any\/} parameter combination, then this indicates significant discordance that cannot be removed. A toy example of such a posterior is shown in \cref{fig:rotate}.
The advantage of building a general dimensional parameterisation-independent prescription to quantify tension is that one can detect discrepancies even if none of the traditional parameters show obvious tension in their marginalised plots.

\subsection{Evidences and Kullback-Leibler divergences}

The numerical evidences and Kullback-Leibler divergences computed from runs produced by \PolyChord{} using the technique described in \cref{sec:ns} are reported in \cref{fig:raw}. 

The first thing to note is that nested sampling does not produce an exact value for the evidence and KL divergence, but instead produces a correlated probability distribution. The correlation is negative, since the dominant error in the evidence estimate is associated with the cumulative Poisson noise in estimating the prior volume contraction at each iteration, and this error contributes equally to both the evidence and KL estimates. Note however that this is advantageous when we wish to compute the $\log S$ ratio, since the error is minimal for the parameters contribution $\log \mathcal{Z} + \mathcal{D}$, as these prior volume errors cancel out to a large extent.

The second observation that should be made is that as we adjust the priors, the log-evidences increase as the normalisation of the prior changes, the Kullback-Leibler divergences decrease since there is less compression between prior and posterior, but the combination $\log \mathcal{Z} + \mathcal{D}$ remains approximately constant.

\subsection{Bayesian model dimensionalities}
The Bayesian model dimensionality for $\Lambda$CDM is detailed for each dataset and prior in \cref{tab:dim}. As this is the first time such quantities have been utilised in a cosmological setting, they are worthy of some discussion.

\begin{figure*}
	\includegraphics[width=\textwidth]{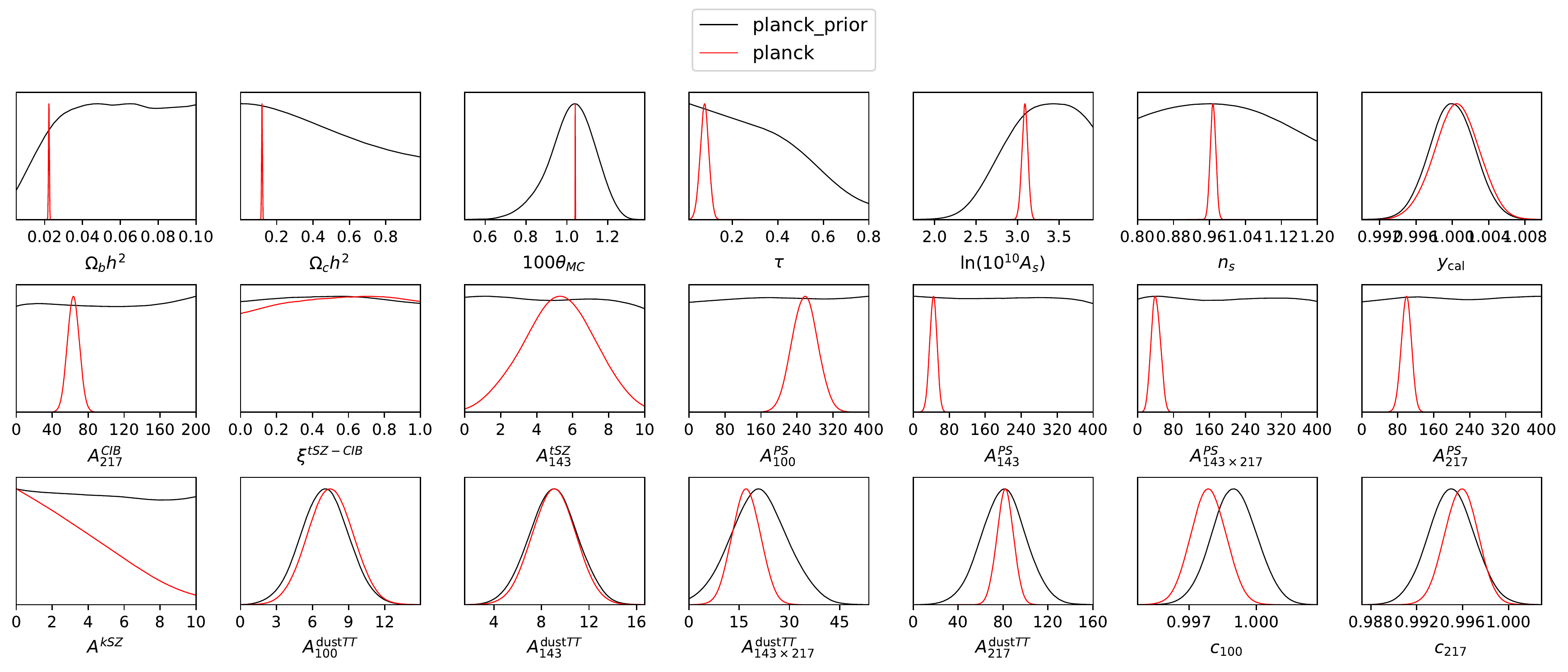}
	\caption{%
        One-dimensional marginalised default prior (black) and \Planck{} posterior (red). The Bayesian model dimensionality of $\tilde{d}_{\Planck}\approx 16$ is reflected by the fact that only a subset of the nuisance parameters are constrained by the data.\label{fig:planck_1d}
	}
\end{figure*}
\begin{figure*}
	\includegraphics[width=\textwidth]{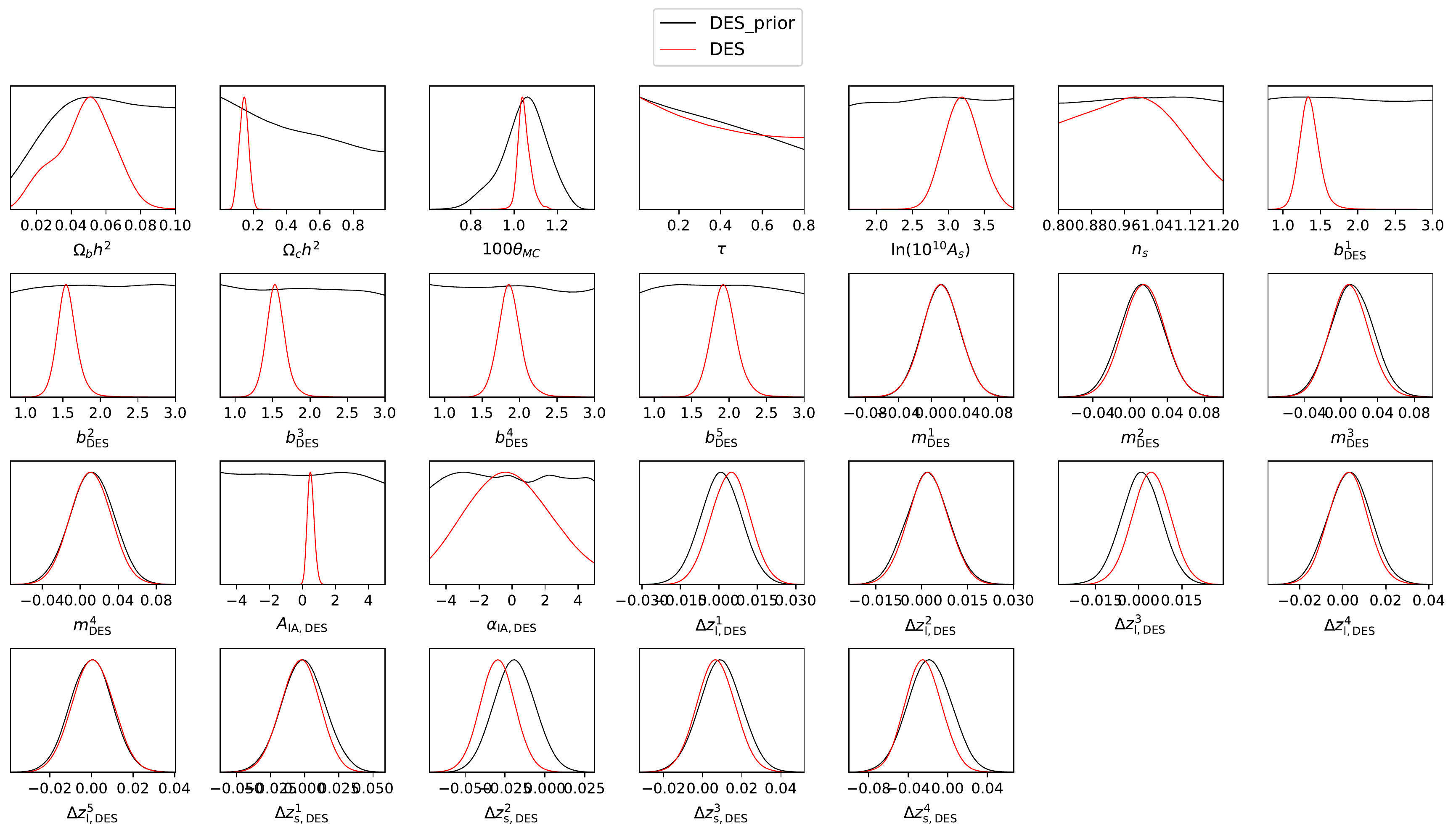}
	\caption{%
        One-dimensional marginalised default prior (black) and DES Y1 posterior (red). The Bayesian model dimensionality of $\tilde{d}_\mathrm{\DES}\approx 14$ is reflected by the fact that only a combination of the cosmological parameters and a subset of the nuisance parameters are constrained by the data.\label{fig:DES_1d}
	}
\end{figure*}

First, the model dimensionality of the \Planck{} dataset remains stable at $\tilde{d}_\Planck{}\approx15$ for all priors. Whilst the $\Planck{}$ 2015 temperature likelihoods nominally have 21 parameters (6 cosmological and 15 nuisance), only a subset of the nuisance parameters are constrained by the data, as can be seen in \cref{fig:planck_1d}. The fact that this dimensionality remains constant for all prior choices is due to the fact that the priors enclose the \Planck{} posterior bulk in all three cases.

Second, in analogy with \Planck{}, the DES Y1 data have a dimensionality of $\tilde{d}_{\mathrm{\DES{}}}\approx11$. As can be seen in \cref{fig:DES_1d}, most of the 20 nuisance parameters and some of the 6 cosmological parameters are unconstrained. Quantifying the dimensionality in this case is made yet harder by the fact that unlike \Planck{}, the DES Y1 survey best constrains a non-trivial combination of the sampled parameters, (e.g.\ $\sigma_8$). It is for this reason that it is essential to have a parameterisation-independent measure of the dimensionality of the constrained parameter space, such as that provided by the Bayesian model dimensionality. Additionally, unlike \Planck{},  for \DES{} there is a slight prior-dependence of the dimensionality for the narrow priors. This can be understood by the fact that the narrow priors cut a little into the \DES{} posterior, effectively rendering some parameters less constrained relative to the wider prior.

This prior dependency is also mirrored in the \SHOES{} and \SDSS{} datasets, although less trivially. For default and medium priors, the dimensionality $\tilde{d}_{\mathrm{S}H_0\mathrm{ES} }=1$ reproduces the correct dimensionality given that the likelihood is only a Gaussian on the Hubble parameter. The fact that this rises to $\tilde{d}_{\mathrm{S}H_0\mathrm{ES} }=2$ for the narrow prior is as a result of a non-trivial degeneracy that emerges for narrow priors in the combination of $(H_0,\Omega_c h^2)$, meaning that the tension constraint of \SHOES{} generates an artificial constraint on $\Omega_c h^2$. The dimensionality of \SDSS{} is yet more complicated, but consistent with the degeneracies between its likelihood and our prior choice.

Finally the combined dimensionalities $\tilde{d}=\tilde{d}_{A} + \tilde{d}_{B}-\tilde{d}_{AB}$ are detailed in the penultimate column of \cref{tab:results}. These show the number of constrained parameters that the datasets have in common, and we can see that \DES{} and \Planck{} share between 1 and $2.5$ constrained parameters depending on the prior chosen. 

In conclusion, there is a rich structure in Bayesian model dimensionalities, and it is our hope that Bayesian model dimensionality becomes more widely used in cosmological inference.

\subsection{Ratios}
We present our key numerical results for the Bayes ratio $R$ and tension probabilities $p$ in \cref{tab:results}.

First, we find that $\log R>0$ for all priors considered for the \SDSS{}+\Planck{} combination, indicating that BAO+RSD datasets are consistent with CMB. More precisely, knowledge of the \SDSS{} dataset boosts our probabilistic confidence in the CMB data by a factor $\sim500$ for the default priors, or $\sim16$ for the narrow priors.  We find that $\log S$ is positive and around zero, with a corresponding tension probability $p\gg 5\%$. One should note that $\log S$ and $p$ are not quite prior-independent since the narrowed priors impinge somewhat on the posterior bulk of the \SDSS{} dataset.

Second for \SHOES{}+\Planck{}, we find that $\log R<0$ for all priors, with our confidence in CMB data dropping in light of knowing the SNe data for all choices of prior, indicating inconsistency. This is also reflected in the tension probabilities, which indicate $p\sim0.3\%$ probability of getting such inconsistency by chance.

Finally, for \DES{} data, the default priors show $R\sim 20$, whilst the narrow priors give $R\sim 0.1$. Under \cref{proposal1}, given that there are some priors which indicate a reduction in confidence in CMB data in light of 3x2 data, we should therefore not regard the datasets as being consistent. Considering the tension statistic, there is a roughly $2\%$ probability of getting such an inconsistency by chance alone. We would therefore consider \DES{} data to be in moderate tension with \Planck{}.  

\subsection{Comparison with the DES analysis}

It should be noted that our conclusion of moderate tension between \DES{} and \Planck{} is in contradiction to that presented in DES Y1. In DES Y1, they compute $R=2.8$, and therefore conclude that there is no tension with CMB data, and hence the datasets are safe to use in conjunction with one another.  Aside from a consideration of the precise meaning of $R$, which is the focus of the first three sections of this paper, there are several issues with their analysis. First, they do not report the errors arising from computing this quantity via nested sampling. Given that they in general use similar settings to ours, it is conceivable that their value of $2.8$ is close to being consistent with $R=1$. Second they use \MultiNest{} to compute this statistic, which renders the value of $R$ that they compute unreliable. Third, they give no consideration to the prior dependency of the $R$ statistic, or to the fact that a small adjustment to their priors would have generated $R<1$. Whilst this dependency is undesirable for some analysts, it should be noted that consistent datasets (e.g.  \SDSS{} and \Planck{}) in general should have $R\gg1$, independent of prior choice. 

\section{Conclusion}\label{sec:conclusion}

In this paper, we examined the Bayes ratio statistic used by DES to quantify the tension between potentially discordant datasets. We provided a novel interpretation of this statistic as a Bayesian quantification of our confidence in our ability to combine the datasets. It represents the factor by which our degree of belief in a dataset is strengthened in light of having incorporated the information provided by another dataset. We explain why this number is prior dependent, and under \cref{proposal1} say that if there is {\em any \/} reasonable prior choice which brings the factor to less than unity, then the datasets should be considered discordant.

For those who mislike the prior dependency of the Bayes ratio, we provide a method of calibrating the statistic using Kullback-Leibler divergences. Inspired by the Gaussian case, \cref{proposal2} provides a Bayesian tension probability, akin to the frequentist $p$-value statistic. As discussed in the introduction, there are several alternative methods for quantifying tensions in the literature, but we claim that this is the only method that preserves all the desiderata of the Bayes ratio, whilst remaining insensitive to prior volume effects.

We applied these new techniques and interpretations to CMB data from \Planck{} combined with the 3x2 data from \DES{}, the BAO+RSD data from \SDSS{} or the SNe data from \SHOES{}. Our technique confirms the consensus view that in comparison with the CMB, there is strong tension with SNe, moderate tension with 3x2 and no tension with BAO+RSD.

We believe that the $R$ statistic is a valuable one for the community to use to compute tension between datasets, but that care must be taken with its interpretation. We hope that these considerations will be taken into account in future \DES{} releases.

\begin{acknowledgements}

    This work was performed using resources provided by the \href{http://www.csd3.cam.ac.uk/}{Cambridge Service for Data Driven Discovery (CSD3)} operated by the University of Cambridge Research Computing Service, provided by Dell EMC and Intel using Tier-2 funding from the Engineering and Physical Sciences Research Council (Capital Grant No.\ EP/P020259/1), and \href{www.dirac.ac.uk}{DiRAC funding from the Science and Technology Facilities Council}.

    This interpretation arose as a result of the KICC conference ``Consistency of Cosmological Datasets: Evidence for new Physics?''\footnote{\href{https://www.ast.cam.ac.uk/meetings/2018/consistency.cosmological.datasets.evidence.new.physics}{https://www.ast.cam.ac.uk/meetings/2018/ \\ consistency.cosmological.datasets.evidence.new.physics}}, to which we are indebted to many participants for extremely fruitful conversations. In particular, we are thankful for discussions with George Efstathiou, Steve Gratton, Ofer Lahav, Mike Hobson and Anthony Lasenby. We would also like to thank Robert Schuhmann and Benjamin Joachimi for help with the Box-Cox transformations, and members of the DES collaboration for useful discussion, especially Scott Dodelson, Marco Raveri, Andresa Campos, and Vivian Miranda. We would also like to thank Daniel Mortlock, for comments on the first version of this paper which led to the use of a superior measure of model dimensionality, and to Andreu Font-Ribera for comments on the final version of the paper. 

W.J.H.\ thanks Gonville \& Caius College for their continuing support via a Research Fellowship. P.L.\ thanks STFC \& UCL for their support via a STFC Consolidated Grant.

\end{acknowledgements}

\bibliographystyle{unsrtnat}
\bibliography{R}

\end{document}